\documentclass[11pt,a4paper]{article}
\usepackage[hyperref]{acl2019}
\usepackage{times}
\usepackage{latexsym}
\usepackage{booktabs} 
\usepackage{lipsum}
\usepackage{tabularx,ragged2e,booktabs}
\usepackage{float}
\usepackage{amssymb}
\usepackage{mathtools}
\usepackage{xcolor}
\usepackage{bbm}
\usepackage{dsfont}
\usepackage{epsfig}
\usepackage{setspace}
\usepackage{subfigure}
\usepackage{hyperref}
\usepackage{microtype}
\usepackage{graphicx}
\usepackage{tabularx}
\usepackage{tabulary}
\usepackage{color}
\usepackage[]{caption}
\usepackage{booktabs} 
\usepackage{lipsum}
\usepackage{tabularx,ragged2e,booktabs}
\usepackage{float}
\usepackage{amssymb}
\usepackage{mathtools}
\usepackage{xcolor}
\usepackage{booktabs}
\usepackage{graphicx}

\usepackage{url}

\usepackage{multirow}

\aclfinalcopy 
\def\aclpaperid{2608} 

\newcommand{\tc}{TC}
\newcommand{\rc}{RC}

\title{A Multi-Task Architecture on Relevance-based Neural Query Translation}

\author{Sheikh Muhammad Sarwar \qquad 
  Hamed Bonab \qquad 
  James Allan \\
  Center for Intelligent Information Retrieval \\
  College of Information and Computer Sciences \\
  University of Massachusetts Amherst \\
 Amherst, MA 01003\\
  \texttt{\{smsarwar, bonab, allan\}@cs.umass.edu} \\
  }

\date{}

\begin{document}

\maketitle

\begin{abstract}
We describe a multi-task learning approach to train a Neural Machine Translation (NMT) model with a Relevance-based Auxiliary Task (RAT) for search query translation. The translation process for Cross-lingual Information Retrieval (CLIR) task is usually treated as a black box and it is performed as an independent step. However, an NMT model trained on sentence-level parallel data is not aware of the vocabulary distribution of the retrieval corpus. We address this problem with our multi-task learning architecture that achieves 16\% improvement over a strong NMT baseline on Italian-English query-document dataset. We show using both quantitative and qualitative analysis that our model generates balanced and precise translations with the regularization effect it achieves from multi-task learning paradigm. 
\end{abstract}

\section{Introduction}





CLIR systems retrieve documents written in a language that is different from search query language \cite{book2010Nie}. The  primary objective of CLIR is to translate or project a query into the language of the document repository \cite{sokokov2013boosting}, which we refer to as Retrieval Corpus (RC). To this end, common CLIR approaches translate search queries using a Machine Translation (MT) model and then use a monolingual IR system to retrieve from \rc. In this process, a translation model is treated as a black box \citep{Sokolov:2014}, and it is usually trained on a sentence level parallel corpus, which we refer to as Translation Corpus (TC).  

We address a pitfall of using existing MT models for query translation \cite{sokokov2013boosting}. An MT model trained on TC does not have any knowledge of RC. In an extreme setting, where there are no common terms between the target side of TC and RC, a well trained and tested translation model would fail because of vocabulary mismatch between the translated query and documents of RC. Assuming a relaxed scenario where some commonality exists between two corpora, a translation model might still perform poorly, favoring terms that are more likely in TC but rare in RC. 
Our hypothesis is that a search query translation model would perform better if a translated query term is likely to appear in the both retrieval and translation corpora, a property we call \emph{balanced translation}.  

To achieve balanced translations, it is desired to construct an MT model that is aware of RC vocabulary. Different types of MT approaches have been adopted for CLIR task, such as dictionary-based MT, rule-based MT, statistical MT etc. \cite{zhou2012}. However, to the best of our knowledge, a neural search query translation approach has yet to be taken by the community. NMT models with attention based encoder-decoder techniques have achieved state-of-the-art performance for several language pairs \cite{bahdanau2014neural}. We propose a multi-task learning NMT architecture that takes RC vocabulary into account by learning Relevance-based Auxiliary Task (RAT). RAT is inspired from two word embedding learning approaches: Relevance-based Word Embedding (RWE) \cite{Zamani:2017} and Continuous Bag of Words (CBOW) embedding \cite{Mikolov:2013}. We show that learning NMT with RAT enables it to generate balanced translation. 

NMT models learn to encode the meaning of a source sentence and decode the meaning to generate words in a target language \cite{LuongPM15}. In the proposed multi-task learning model, RAT shares the decoder embedding and final representation layer with NMT. Our architecture answers the following question: \emph{In the decoding stage, can we restrict an NMT model so that it does not only generate terms that are highly likely in TC?}. We show that training a strong baseline NMT with RAT roughly achieves 16\% improvement over the baseline. Using a qualitative analysis, we further show that RAT works as a regularizer and prohibits NMT to overfit to TC vocabulary.  




\section{Balanced Translation Approach}
We train NMT with RAT to achieve better query translations. We improve a recently proposed NMT baseline, Transformer, that achieves state-of-the-art results for sentence pairs in some languages \cite{vaswani}. We discuss Transformer, RAT, and our multi-task learning architecture that achieves balanced translation. 

\subsection{NMT and Transformer} 
In principle, we could adopt any NMT and combine it with RAT. An NMT system directly models the conditional probability $P(t_i \vert s_i)$ of translating a source sentence, $s_i = s_i^1,\ldots , s_i^n$, to a target sentence $t_i = t_i^1, \ldots , t_i^n$. A basic form of NMT comprises two components: (a) an \emph{encoder} that computes the representations or meaning of $s_i$ and (b) a \emph{decoder} that generates one target word at a time. State-of-the-art NMT models have an attention component that ``searches for a set of positions in a source sentence where the most relevant information is concentrated'' \cite{bahdanau2014neural}.   

For this study, we use a state-of-the-art NMT model, Transformer \cite{vaswani}, that uses positional encoding and self attention mechanism to achieve three benefits over the existing convolutional or recurrent neural network based models: (a) reduced computational complexity of each layer, (b) parallel computation, and (c) path length between long-range dependencies.

\subsection{Relevance-based Auxiliary Task (RAT)}
We define RAT a variant of word embedding task \cite{Mikolov:2013}. Word embedding approaches learn high dimensional dense representations for words and their objective functions aim to capture contextual information around a word. \citet{Zamani:2017} proposed a model that learns word vectors by predicting words in relevant documents retrieved against a search query. We follow the same idea but use a simpler learning approach that is suitable for our task. They tried to predict words from the relevance model \cite{Lavrenko2003} computed from a query, which does not work for our task because the connection between a query and ranked sentences falls rapidly after the top one (see below). 

We consider two data sources for learning NMT and RAT jointly. The first one is a sentence-level parallel corpus, which we refer to as translation corpus, $\tc = \{(s_i, t_i); i= 1, 2, \ldots m\}$. The second one is the retrieval corpus, which is a collection of $k$ documents $\rc = \{D_1, D_2, \ldots D_k\}$ in the same language as $t_i$. Our word-embedding approach takes each $t_i \in \tc$, uses it as a query to retrieve the top document $D_i^{top}$. After that we obtain $t_i'$ by concatenating $t_i$ with $D_i^{top}$ and randomly shuffling the words in the combined sequence. We then augment $\tc$ using $t_i^\prime $ and obtain a dataset, $TC' =  \{(s_i, t_i, t_i^\prime); i= 1, 2, \ldots m\}$. We use $t_i'$ to learn a continuous bag of words (CBOW) embedding as proposed by \citet{Mikolov:2013}. This learning component shares two layers with the NMT model. The goal is to expose the retrieval corpus' vocabulary to the NMT model. We discuss layer sharing in the next section.  

We select the single top document retrieved against a sentence $t_i$ because a sentence is a weak representation of information need. As a result, documents at lower ranks show heavy shift from the context of the sentence query. We verified this by observing that a relevance model constructed from the top $k$ documents does not perform well in this setting. We thus deviate from the relevance model based approach taken by \citet{Zamani:2017} and learn over the random shuffling of $t_i$ and a single document. Random shuffling has shown reasonable effectiveness for word embedding construction for comparable corpus \cite{vulic2015monolingual}. 



\begin{figure}[t]
    \centering
    \includegraphics[width=\linewidth ]{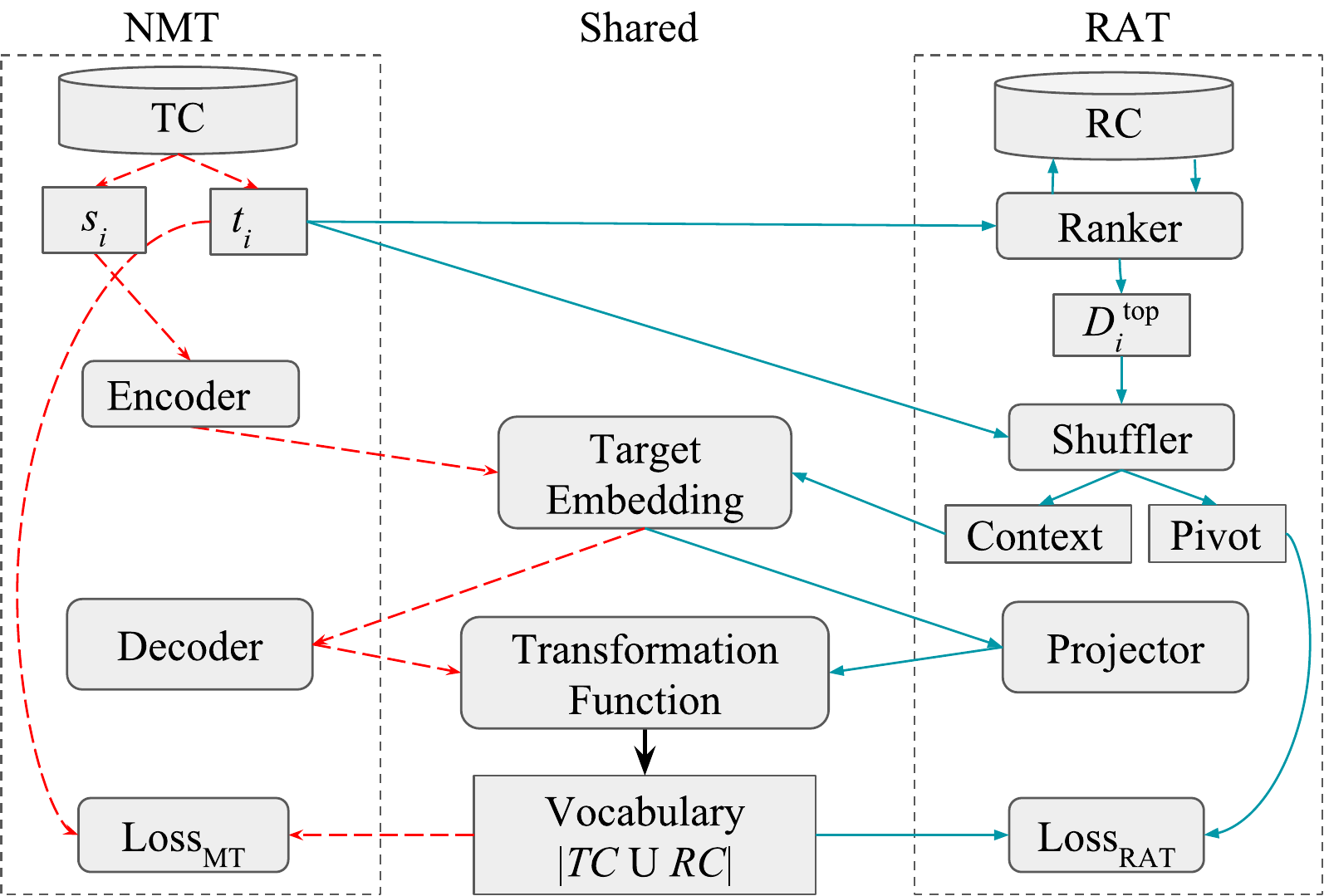}
    \caption{The architecture of our multi-task NMT. Note that, rectangles indicate data sources and rectangles with rounded corners indicate functions or layers.}
    \label{fig:mtask_nmt}
\end{figure}

\subsection{Multi-task NMT Architecture}
Our balanced translation architecture is presented in Figure~\ref{fig:mtask_nmt}. This architecture is NMT-model agnostic as we only propose to share two layers common to most NMTs: the trainable target embedding layer and the transformation function \cite{LuongPM15} that outputs a probability distribution over the union of the vocabulary of TC and RC. Hence, the size of the vocabulary, $\vert \rc \cup \tc \vert$, is much larger compared to \tc{} and it enables the model to access \rc. In order to show task sharing clearly we placed two shared layers between NMT and RAT in Figure \ref{fig:mtask_nmt}. We also show the two different paths taken by two different tasks at training time: the NMT path in shown with red arrows while the RAT path is shown in green arrows. 

On NMT path training loss is computed as the sum of term-wise softmax with cross-entropy loss of the predicted translation and the human translation and it summed over a batch of sentence pairs, $\mathcal{L}_{NMT} = \sum_{(s_i, t_i) \in T} \sum_{j=1}^{\vert t_i \vert} -\log P(t_i^j\vert t_i^{<j}, s_i)$. We also use a similar loss function to train word embedding over a set of context ($ctx$) and pivot ($pvt$) pairs formed using $t_i$ as query to retrieve $D_i^{top}$ using Query Likelihood (QL) ranker, $\mathcal{L}_{WE} = \alpha \underset{(ctx, pvt)} {\sum} -\log P(pvt \mid ctx)$. This objective is similar to CBOW word embedding as context is used to predict pivot word 
Here, we use a scaling factor $\alpha$, to have a balance between the gradients from the NMT loss and RAT loss. For RAT, the context is drawn from a context window following \citet{Mikolov:2013}.  

In the figure, $(s_i,t_i) \in \tc$ and $D_i^{top}$ represents the top document retrieved against $t_i$. The shuffler component shuffles $t_i$ and $D_i^{top}$ and creates (context, pivot) pairs. After that those data points are passed through a fully connected linear projection layer and eventually to the transformation function. Intuitively, the word embedding task is similar to NMT as it tries to assign a large probability mass to a target word given a context. However, it enables the transformation function and decoding layer to assign probability mass not only to terms from \tc, but also to terms from \rc. This implicitly prohibits NMT to overfit and provides a regularization effect. A similar technique was proposed by \citet{nmtwemb} to handle out-of-vocabulary or less frequent words for NMT. For these terms they enabled the transformation (also called the softmax cross-entropy layer) to fairly distribute probability mass among similar words. In contrast, we focus on \emph{relevant} terms rather than similar terms.

\section{Experiments and Results} 
\paragraph{Data.} We experiment on two language pairs: \{Italian, Finnish\} $\rightarrow$ English. Topics and relevance judgments are obtained from the Cross-Language Evaluation Forum (CLEF) 2000-2003 campaigns for bilingual ad-hoc retrieval tracks\footnote{catalog.elra.info/en-us/repository/browse/ELRA-E0008/}. The Italian and French topics are human translations of a set of two hundred English topics. Our retrieval corpus is the Los Angeles Times (LAT94) comprising over 113k news articles.

Topics without any relevant documents on LAT94 are excluded resulting in 151 topics for both Italian and Finnish language. Among the 151 topics in our dataset, we randomly selected 50 queries for validation and 101 queries for test. In the CLEF literature, queries are constructed from either the \textit{title} field or a concatenation of \textit{title} and \textit{description} fields of the topic sets. Following \citet{vulic2015monolingual}, we work on the longer queries. 

For \tc~we use Europarl v7 sentence-aligned corpus \cite{koehn2005europarl}. TC statistics in Table \ref{tab:parallelcorpus} indicates that we had around two million sentence pairs for each language pairs.  
\begin{table}[!h]
    \centering
    {
        \small
    	\tabcolsep=0.19cm
        \begin{tabular}{llrrr} \toprule
            Lang. Pair  & Resource &  \#Inst. & $|V_F|$ & $|V_E|$\\ \midrule
            \multirow{1}{4em}{Ita-Eng}   & Europarl   & 1,894,217 & 146,036 & 77,441\\ 
                                          \midrule
            \multirow{1}{4em}{Fin-Eng}   & Europarl   & 1,905,683 & 637,902 & 75,851\\ 
            \bottomrule
        \end{tabular}
    }
    \vspace{-0.3cm}
    \caption{Statistics of resources used for training. $|V_F|$ and $|V_E|$ are the vocabulary size for the source language and the target English language, respectively.  }
    \label{tab:parallelcorpus}
\end{table}

\paragraph{Text Pre-processing.} 
For having text consistency across \tc~ and \rc, we apply the following pre-processing steps. Characters are normalized by mapping diacritic characters to the corresponding unmarked characters and lower-casing. We remove non-alphabetic, non-printable, and punctuation characters from each word. The NLTK library \cite{bird2004nltk} is used for tokenization and stop-word removal. No stemming is performed. 

\paragraph{Retrieval.} For ranking documents, after query translation, we use the Galago's implementation\footnote{https://www.lemurproject.org/galago.php} of query likelihood using Dirichlet smoothing \cite{zhai2004study} with default parameters.

\paragraph{Training Technique.} Before applying multi-tasking we train the transformer to obtain a reasonable MAP on the Val set. Then we spawn our multi-task transformer from that point, also continuing to train the transformer. We use an early stopping criterion to stop both the models, and evaluate performance on the test set. For NMT training we use Stochastic Gradient Descent (SGD) with Adam Optimizer and learning rate of 0.01. We found that a learning rate of $10^{-5}$ with the same optimizer works well for the word embedding loss minimization. From a training batch (we use dynamic size training batches), more data points are actually created for the word embedding task because of large number of (context, pivot) pairs. We allow the gradients from word embedding loss to pass through the multi-tasking model at first, and then apply NMT loss. Setting a lower learning rate for the word embedding optimizer, and $\alpha  = 0.1 $ allows the NMT gradient updates to be competitive. 

\paragraph{Evaluation.}  Given that in CLIR the primary goal is to get a better ranked list of documents against a translated query, we only report Mean Average Precision (MAP).

\subsection{Results and Analysis}
Table \ref{tab:map_comparison} shows the effectiveness of our model (multi-task transformer) over the baseline transformer \cite{vaswani}. Our model achieves significant performance gains in the test sets over the baseline for both Italian and Finnish query translation. The overall low MAP for NMT can possibly be improved with larger TC. Moreover, our model validation approach requires access to RC index, and it slows down overall training process. Hence, we could not train our model for a large number of epochs - it may be another cause of the low performance.  
\begin{table}[t]
\resizebox{\linewidth}{!}{%
\begin{tabular}{@{}lccll@{}}
\toprule
                       & \multicolumn{2}{l}{Italian $\rightarrow$ English} & \multicolumn{2}{l}{Finnish $\rightarrow$ English}             \\ 
Models                 & Val              & Test              & \multicolumn{1}{c}{Val} & \multicolumn{1}{c}{Test} \\ \midrule
Transformer            & 0.192               & 0.179              &                 \textbf{0.127}        &   0.077                        \\ 
Our model & \textbf{0.230}               & \textbf{0.211}              &            0.126             &  \textbf{0.097}                        \\ \bottomrule
\end{tabular}%
}
	\vspace{-0.3cm}
\caption{Results for ranking with query translation models, in terms of MAP.}
\label{tab:map_comparison}
\end{table}

\begin{figure}[t]
    \centering
    \includegraphics[width=\linewidth]{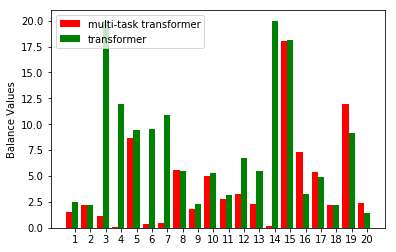}
    \caption{Balance values of a sample of val queries}
    \label{fig:qval}
\end{figure}

\begin{figure}[t]
    \centering
    \includegraphics[width=\linewidth]{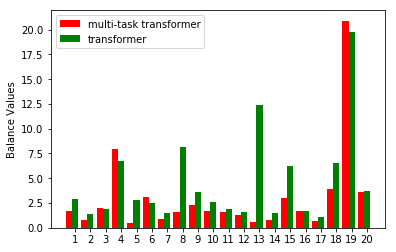}
    \caption{Balance values of a sample of test queries}
    \label{fig:qtest}
\end{figure}

\paragraph{Balance of Translations.}
We want to show that translation terms generated by our multi-task transformer  are roughly equally likely to be seen in the Europarl corpus (\tc) or the CLEF corpus (\rc). Given a translation term $t$, we compute the ratio of the probability of seeing $t$ in \tc{} and \rc, $\frac{P_{\tc}(t)}{P_{\rc}(t)}$. Here, $P_{\tc}(t) = \frac{count_{\tc}(t)}{\sum_{t \in {\tc}} count_{\tc}(t)}$ and $P_{\rc}(t)$ is calculated similarly. Given a query $q_i$ and its translation $T_m(q_i)$ provided by model $m$, we calculate the balance of $m$, $B(T_m(q_i)) = \frac{\sum_{t \in T_m(q)} \frac{P_{\tc}(t)}{P_{\rc}(t)}}{\mid T_m(q) \mid}$. If $B(T_m(q_i))$ is close to 1, the translation terms are as likely in \tc{} as in \rc. Figure \ref{fig:qval} shows the balance values for transformer and our model for a random sample of 20 queries from the validation set of Italian queries, respectively. Figure \ref{fig:qtest} shows the balance values for transformer and our model for a random sample of 20 queries from the test set of Italian queries, respectively. It is evident that our model achieves better balance compared to baseline transformer, except for a very few cases.

\paragraph{Precision and Recall of Translations.} 
Given a query $Q$ , consider $Q^\prime = \{q^\prime_1, q^\prime_1, \ldots , q^\prime_p\}$ as the set of terms from human translation of $Q$ and $Q^M = \{q^M_1, q^M_2, \ldots ,q^M_q\}$ as the set of translation terms generated by model $M$. We define $P_M(Q) = \frac{Q^M \cap Q^\prime}{\vert Q^M \vert}$ and $R_M(Q) = \frac{Q^M \cap Q^\prime}{\vert Q^\prime \vert}$ as precision and recall of $Q$ for model $M$. In Table \ref{tab:translation_quality_comparison}, we report average precision and recall for both transformer and our model across our train and validation query set over two language pairs. Our model generates precise translation, \emph{i.e.} it avoids terms that might be useless or even harmful for retrieval. Generally, from our observation, avoided terms are highly likely terms from TC and they are generated because of translation model overfitting. Our model achieves a regularization effect through an auxiliary task. This confirms results from existing multi-tasking literature \cite{ruder2017overview}.  

To explore translation quality, consider pair of sample translations provided by two models. For example, against an Italian query, \emph{medaglia oro super vinse medaglia oro super olimpiadi invernali lillehammer}, translated term set from our model is \{gold, coin, super, free, harmonising, won, winter, olympics\}, while transformer output is \{olympic, gold, one, coin, super, years, won, parliament, also, two, winter\}. Term set from human translation is: \{super, gold, medal, won, lillehammer, olypmic, winter, games\}. Transformer comes up with terms like \emph{parliament}, \emph{also}, \emph{two} and \emph{years} that never appears in human translation. We found that these terms are very likely in Europarl and rare in CLEF. Our model also generates terms such as \emph{harmonising}, \emph{free}, \emph{olympics} that not generated by transformer. However, we found that these terms are equally likely in Europarl and CLEF.    

\begin{table}[t]
\resizebox{\linewidth}{!}{%
\begin{tabular}{@{}p{5 em}ccll@{}}
\toprule
                       & \multicolumn{2}{c}{Italian $\rightarrow$ English} & \multicolumn{2}{c}{Finnish $\rightarrow$ English}             \\ 
Models                 & Val              & Test              & \multicolumn{1}{c}{Val} & \multicolumn{1}{c}{Test} \\ \midrule
Transformer &  (0.44, \textbf{0.45})    &        (0.43, \textbf{0.46})       &      (0.24, 0.23)   & (0.25, \textbf{0.26})                      \\ 
Our model            &      (\textbf{0.62}, \textbf{0.45})           &      (\textbf{0.57}, 0.41)         &  (\textbf{0.31}, \textbf{0.25})                       &    (\textbf{0.30}, 0.24)                      \\\bottomrule

\end{tabular}%
}
\vspace{-0.3cm}
\caption{Average precision and recall of translated queries, respectively reported in tuples.}
\label{tab:translation_quality_comparison}
\end{table}


\section{Conclusion}
We present a multi-task learning architecture to learn NMT for search query translation. As the motivating task is CLIR, we evaluated the ranking effectiveness of our proposed architecture. We used sentences from the target side of the parallel corpus as queries to retrieve relevant document and use terms from those documents to train a word embedding model along with NMT. One big challenge in this landscape is to sample meaningful queries from sentences as sentences do not directly convey information need. In the future, we hope to learn models that are able to sample search queries or information needs from sentences and use the output of that model to get relevant documents.  



\section*{Acknowledgments}
This work was supported in part by the Center for Intelligent Information Retrieval and in part by the Air Force Research Laboratory (AFRL) and IARPA under contract \#FA8650-17-C-9118 under subcontract \#14775 from Raytheon BBN Technologies Corporation. Any opinions, findings and conclusions or recommendations expressed in this material are those of the authors and do not necessarily reflect those of the sponsor.

\bibliographystyle{acl_natbib}
\bibliography{acl2019}

\appendix

\section{Loss Function and
Validation Performance Analysis}
\label{sec:loss_function_analysis}
\begin{figure}[!htpb]
    \centering
    \includegraphics[width=0.9\linewidth]{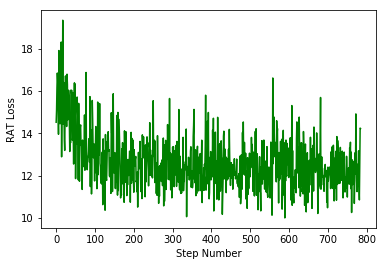}
    \caption{RAT loss of our model on Italian-English training data}
    \label{fig:rat_loss}
\end{figure}

\begin{figure}[!htpb]
    \centering
    \includegraphics[width=0.9\linewidth]{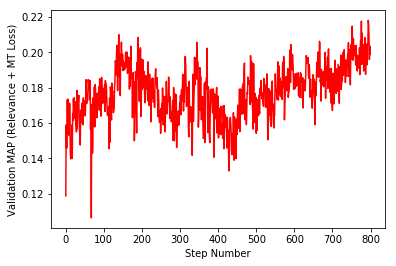}    \caption{Validation performance of our model on Italian-English validation data}
    \label{fig:val_our}
\end{figure}

\begin{figure}[!htpb]
    \centering
    \includegraphics[width=0.9\linewidth]{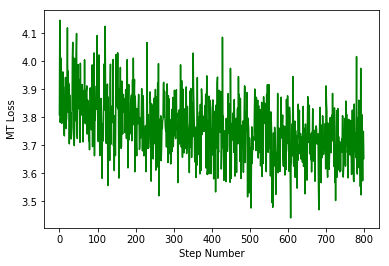}
    \caption{NMT loss of our model on Italian-English training data}
    \label{fig:mt_loss}
\end{figure}

\begin{figure}[!htpb]
    \centering
    \includegraphics[width=0.9\linewidth]{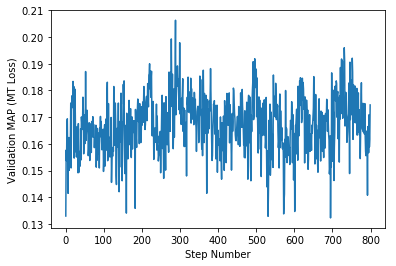}
    \caption{Validation set performance of Transformer on Italian-English training data}
    \label{fig:val_transformer}
\end{figure}

We show the loss function analysis of transformer and our model. Figure \ref{fig:val_transformer} shows the validation performance of transformer against global training steps. Figure \ref{fig:val_our} show the validation performance of our model for the same number of global steps. Figure \ref{fig:mt_loss} shows that NMT loss is going down with the number of steps, while Figure \ref{fig:rat_loss} shows the degradation of the loss of our proposed RAT task. 



\end{document}